\documentclass[10pt,journal,twoside]{IEEEtran}
\usepackage{multirow}
\usepackage{diagbox}
\usepackage{amssymb}
\usepackage[dvips]{graphicx}
\usepackage{amsmath}
\usepackage{amsthm}
\usepackage{latexsym,bm}
\usepackage{color}
\usepackage{subfigure}
\usepackage{longtable}

\usepackage{cite}
\usepackage{enumerate}
\usepackage{arydshln}
\usepackage{slashbox}
\usepackage{algorithmic}
\usepackage{float}
\usepackage{booktabs}
\usepackage{subfigure}
\usepackage{stfloats}
\usepackage[table,xcdraw]{xcolor}
\usepackage{graphicx}
\usepackage{epstopdf}
\usepackage{bm}

\makeatletter
\def\widebar{\accentset{{\cc@style\underline{\mskip10mu}}}}
\def\Widebar{\accentset{{\cc@style\underline{\mskip8mu}}}}
\makeatother

\theoremstyle{plain}

\theoremstyle{definition}
\theoremstyle{definition}

\begin{document}
\title{\vspace{-4mm}Design of a New CIM-DCSK-Based Ambient\\Backscatter Communication System}
\author{Ruipeng~Yang, Yi~Fang, \IEEEmembership{Senior~Member, IEEE}, Pingping~Chen, \IEEEmembership{Senior~Member, IEEE},
and Huan~Ma, \IEEEmembership{Member, IEEE}\vspace{-11mm}%
\thanks{R.~Yang, and Y.~Fang are with the School of Information Engineering, Guangdong University of Technology, Guangzhou 510006, China, (email: 1112103004@mail2.gdut.edu.cn; fangyi@gdut.edu.cn).}%
\thanks{H.~Ma is with the Electronics and Information Engineering Department, Zhaoqing University, Zhaoqing 526061, China, (email: mh-zs@163.com).}%
\thanks{P.~Chen is with the Department of Electronic Information, Fuzhou University, Fuzhou 350116, China (e-mail: ppchen.xm@gmail.com).}%
}\vspace{-20mm}%
%\vspace{-20mm}
\maketitle
%\vspace{-25mm}
\begin{abstract}
To improve the data rate in differential chaos shift keying (DCSK) based ambient backscatter communication (AmBC) system, we propose a new AmBC system based on code index modulation (CIM), referred to as CIM-DCSK-AmBC system. In the proposed system, the CIM-DCSK signal transmitted in the direct link is used as the radio frequency source of the backscatter link. The signal format in the backscatter link is designed to increase the data rate as well as eliminate the interference of the direct link signal. As such, the direct link signal and the backscatter link signal can be received and demodulated simultaneously. Moreover, we derive and validate the theoretical bit error rate (BER) expressions of the CIM-DCSK-AmBC system over multipath Rayleigh fading channels. Regarding the short reference DCSK-based AmBC (SR-DCSK-AmBC) system as a benchmark system, numerical results reveal that the CIM-DCSK-AmBC system can achieve better BER performance in the direct link and higher throughput in the backscatter link than the benchmark system.
\end{abstract}
\vspace{-0.1cm}
\vspace{-0.45cm}
\section{Introduction}
\IEEEPARstart{D}{ifferential} chaos shift keying (DCSK) system is a current research hot spot in spread-spectrum communication owing to its outstanding performance over multipath fading channels. However, because DCSK is a transmitted-reference (TR) communication system, half of the symbol period is used as a reference, which results in a low data rate.

To tackle this problem, code-shifted DCSK \cite{6220867} and multi-carrier DCSK (MC-DCSK) \cite{6560492} have been presented with increased data rates, avoiding the use of RF delay lines. Quadrature CSK \cite{972858} has extended the DCSK system to the non-binary system. Index modulation \cite{9866575} has become another high-data-rate technique applied to the DCSK system. By employing Walsh codes to transmit additional information, CIM-DCSK systems have been proposed in \cite{8290668,9277910,9475491}. A permutation-matrix-based permutation index DCSK system has been introduced in \cite{8110728} achieving multi-user high-data-rate transmission with enhanced system security. In addition, a hybrid index modulation DCSK is proposed in \cite{9761226}, which integrates carrier index and carrier number index into MC-DCSK. Moreover, multidimensional index modulation has been introduced in \cite{9841425}.

Recently, ambient backscatter communication (AmBC) has become another promising technique to increase data rates. Using the direct link signal as its energy source, the backscatter link signal in the AmBC system transmits additional symbols without adding power-hungry RF components. Based on the above advantage, an SR-DCSK-based ambient backscatter communication (SR-DCSK-AmBC) system has been proposed in \cite{9525461}. In this system, SR-DCSK is used for the direct link modulation, and a novel signal format has been designed for the backscatter link to eliminate the interference from the direct link signal.

We notice that the repeat chaotic sequence in the direct link signal of SR-DCSK-AmBC is indispensable to eliminate the interference in the system. This salient feature is naturally suitable for combining with CIM-DCSK to further increase the data rate. Thus, we propose a new CIM-DCSK-AmBC system in this paper. In the proposed system, the Walsh codes used in the direct link signal are intelligently selected to adapt to the signal format in the backscatter link, thus the receiver can effectively eliminate the interference between them. In addition, the backscatter link signal is designed to transmit more information bits than the SR-DCSK-AmBC system, while maintaining the anti-direct-link interference capability. The theoretical bit error rate (BER) expressions of the proposed CIM-DCSK-AmBC system are derived over multipath Rayleigh fading channel. Numerical results reveal that the CIM-DCSK-AmBC system provides better BER and throughput than the SR-DCSK-AmBC system and the theoretical BER expressions provide an accurate estimation.
\vspace{-0.45cm}
\section{System Model}
\begin{figure}[tbp]
\center
\vspace{-0.0cm}
\includegraphics[width=2.5 in]{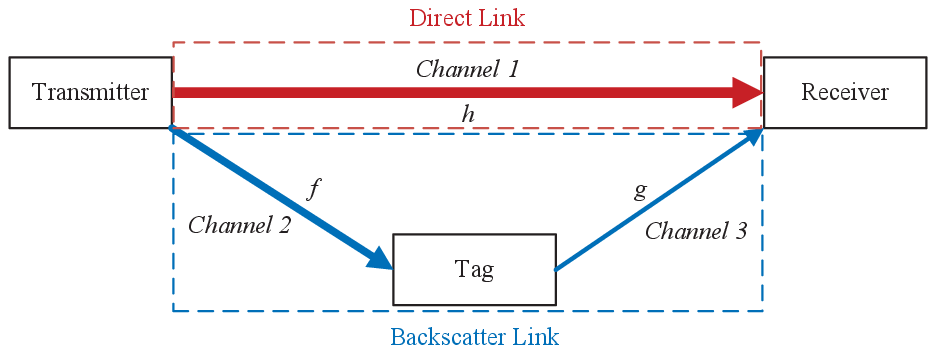}
\vspace{-0.3cm}
\caption{System model of an ambient backscatter communication.}
\label{Model}
\vspace{-0.7cm}
\end{figure}
The model of an ambient backscatter communication system is shown in Fig.~\ref{Model}. As shown in this figure, the channel coefficient in the direct link is denoted by $h$. The two channel coefficients in the backscatter link are denoted by $f$ and $g$, respectively. All channel coefficients are assumed to follow Rayleigh distributions and be independent of one another \cite{9970371}. The direct link signal is transmitted to the receiver through channel $1$ . The backscatter link signal first passes through channel $2$, then is reflected and modulated by the tag. At last, the backscatter link signal is transmitted to the receiver through channel $3$.

\begin{figure}[tbp]
\center
\vspace{-0.0cm}
\includegraphics[width=3.6 in]{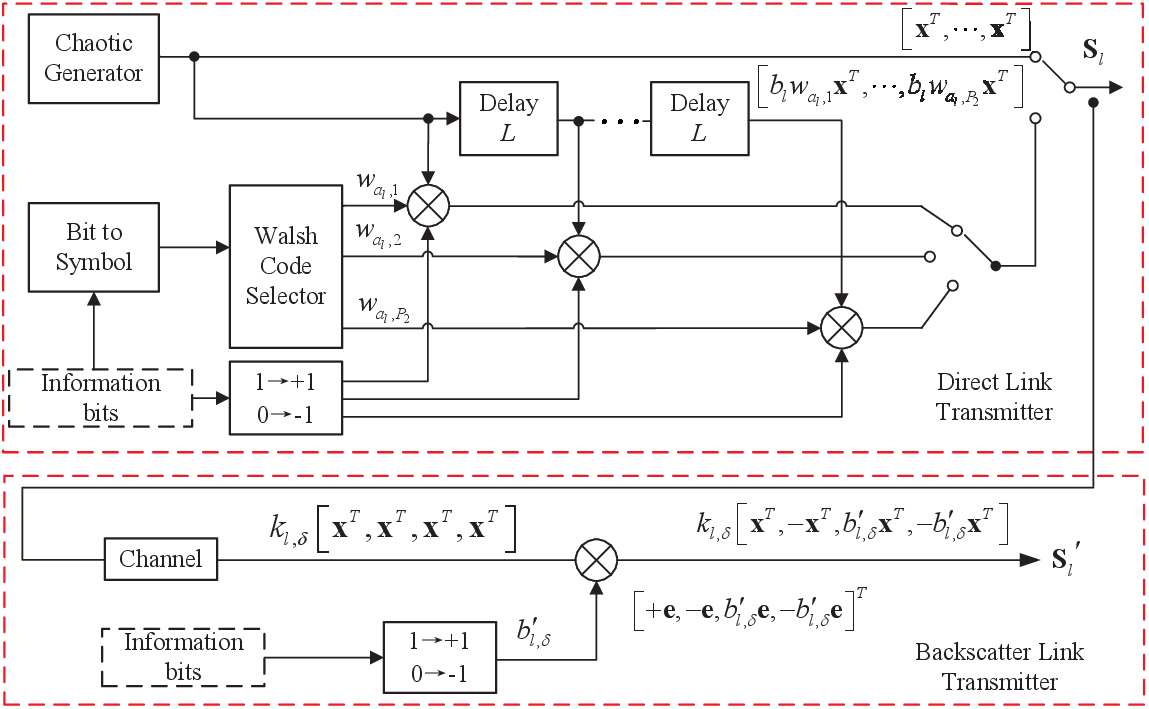}
\vspace{-0.6cm}
\caption{Block diagram of a CIM-DCSK-AmBC transmitter.}
\label{Transmitter}
\vspace{-0.7cm}
\end{figure}

The block diagram of the CIM-DCSK-AmBC transmitter and receiver is shown in Fig.~\ref{Transmitter}, where $\textbf{e}=[1,1,\cdots,1]$ is a length-$L$ vector. The transmitted signal of the $l^{th}$ symbol in the direct link is given by
\begin{equation}
\textbf{S}_l=\left[ \begin{matrix} \underbrace{ \textbf{x}^T,\cdots,\textbf{x}^T, } & \underbrace{b_lw_{a_{l,1}}\textbf{x}^T,\cdots,b_lw_{a_{l,P_2}}\textbf{x}^T}
\\ \text{reference of } P_1\text{ replicas} & \text{information bearing} \end{matrix} \right],
\end{equation}
where $ P_1,P_2 $ is divisible by $4$, $\textbf{x}$ is a length-$L$ chaotic sequence, $b_l$ is the modulated symbol in the direct link of the CIM-DCSK-AmBC system, and $w_{a_l}$ is the $a_l^{th}$ length-$P_2$ Walsh code.
Note that $M=P_2/4$ different Walsh codes are used in the transmitted signal and $a_l\in\left\{1,5,\cdots,1+4(M-1)\right\}$, which are different from the conventional CIM-DCSK. $m_c=\log_2(M)$ index bits can be transmitted in each symbol $\textbf{S}_l$. Thus, ${(a_l-1)}$ is divisible by $4$, every $4$ consecutive elements of the length-$P_2$ Walsh code $\{ w_{a_{l,p_2}},w_{a_{l,p_2+1}},w_{a_{l,p_2+2}},w_{a_{l,p_2+3}} \}$ are the same, where $p_2\in\left\{ 1,5,\cdots,1+(P_2-4) \right\}$. Then, $\textbf{S}_l$ can be rewritten as
\begin{equation}
\textbf{S}_l=\left[  \textbf{s}_{l,1},\textbf{s}_{l,2}, \cdots ,  \textbf{s}_{l,(P_1+P_2)/4} \right],
\end{equation}
where $\textbf{s}_{l,\delta}= k_{l,\delta}[\textbf{x}^T, \textbf{x}^T, \textbf{x}^T, \textbf{x}^T]$, $k_{l,\delta}\in\left\{ -1,+1 \right\}$ is the sign of the direct link symbol in the CIM-DCSK-AmBC system, $\delta \in \left\{  1,2,...,(P_1+P_2)/4 \right\}$. To be more specific, if $\textbf{s}_{l,\delta}$ is a part of the reference of $\textbf{S}_l$, $k_{\delta}=1$. If $\textbf{s}_{l,\delta}$ is a part of the information bearing signal of $\textbf{S}_l$, $k_{l,\delta}=b_lw_{a_{l,p_2}}$, where $p=4\delta-P_4$.

We define $\textbf{s}'_{l,\delta}$ as the signal reflected from $\textbf{s}_{l,\delta}$ by the tag, which can be expressed as
\begin{equation}
\textbf{s}'_{l,\delta}= k_{l,\delta} \left[ \begin{matrix} \underbrace{ \textbf{x}^T,-\textbf{x}^T }, & \underbrace{ b'_{\delta}\textbf{x}^T,-b'_{\delta}\textbf{x}^T }
\\ \text{reference} & \text{information bearing} \end{matrix} \right],
\end{equation}
where $b'_{l,\delta}\in\left\{-1,+1\right\}$ is the CIM-DCSK-AmBC modulated symbol in the backscatter link. Thus, $\textbf{s}_{l,\delta}$ and $\textbf{s}'_{l,\delta}$ are mutually orthogonal. We define $\textbf{S}'_{l}$ as the signal reflected from $\textbf{S}_{l}$ by the tag
\begin{equation}
\textbf{S}'_{l}=\left[  \textbf{s}'_{l,1},\textbf{s}'_{l,2}, \cdots ,  \textbf{s}'_{l,(P_1+P_2)/4} \right],
\end{equation}
Thus, $\textbf{S}'_{l}$ is orthogonal to $\textbf{S}_l$. As a result, the interference between the direct link and the backscatter link can be eliminated well because of the orthogonality of the signals.

{\em Remark: In the proposed system, Walsh codes are intelligently selected, i.e., only $M=P_2/4$ length-$P_2$ Walsh codes are used to ensure the orthogonality between $\textbf{S}_{l}$ and $\textbf{S}'_l$. Therefore, $\textbf{S}'_{l}$ can be removed from the decision variable $I_n$ in the direct link. Similarly, $\textbf{S}_{l}$ can be removed from the decision variable $D_{l,\delta}$ in the backscatter link. Moreover, the remaining $3P_2/4$ length-$P_2$ Walsh codes are not used in the proposed system, because they cannot ensure the orthogonality between $\textbf{S}_{l}$ and $\textbf{S}'_l$. In this case, the backscatter link signal $\textbf{S}'_l$ can be masked by $\textbf{S}_{l}$. Besides, for the signal $\textbf{S}_{l}$, the signal $\textbf{S}'_l$ is a non-negligible interference.}

\begin{figure}[tbp]
\center
\vspace{-0.0cm}
\includegraphics[width=3.6 in]{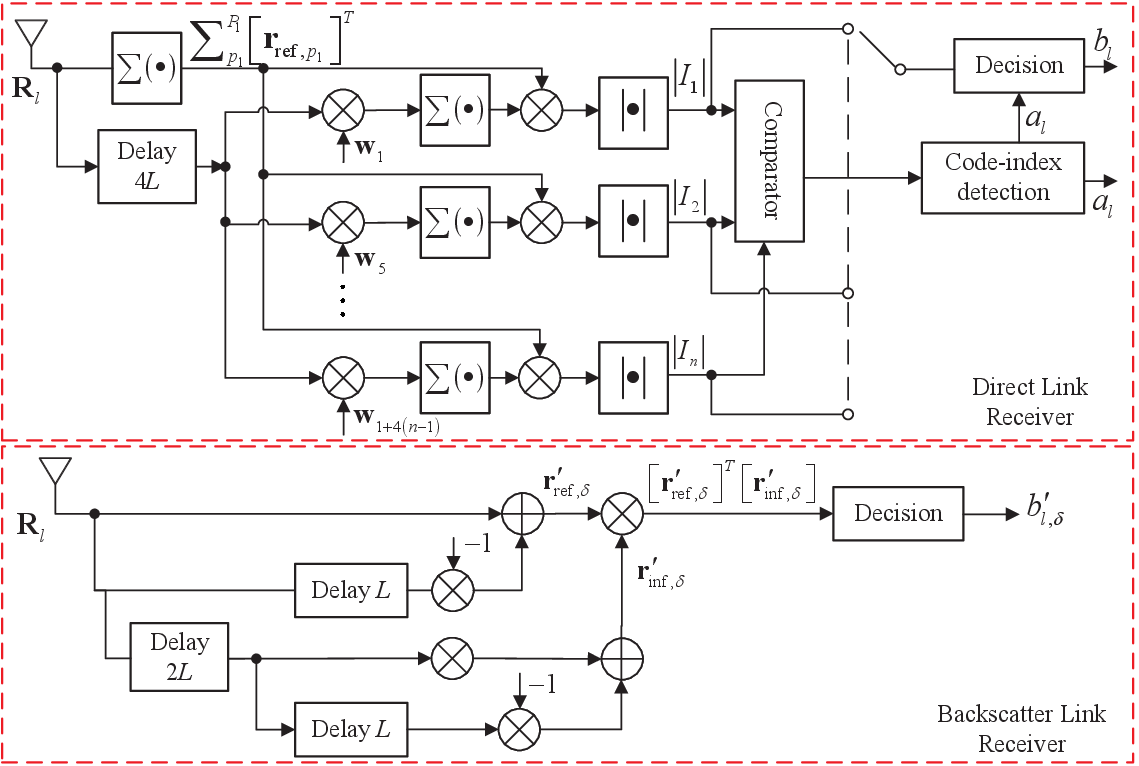}
\vspace{-0.6cm}
\caption{Block diagram of a CIM-DCSK-AmBC receiver.}
\label{Receiver}
\vspace{-0.7cm}
\end{figure}

The block diagram of the CIM-DCSK-AmBC receiver is shown in Fig.~\ref{Receiver}. Both the transmitted signals $\textbf{S}_l$ in the direct link and $\textbf{S}'_l$ in the backscatter link can be received at the same time, as follows
\begin{equation}
 \textbf{S}_l+\textbf{S}'_l+\textbf{n}_l=\left[ r_1,\cdots,r_{(P_1+P_2)L} \right]^T,
\end{equation}

In the direct link, $\textbf{R}_l$ can be presented as
\begin{equation}
 \textbf{R}_l=\left[ \textbf{r}_{\text{ref},1},\cdots,\textbf{r}_{\text{ref},P_1},
 \textbf{r}_{\text{inf},1},\cdots,\textbf{r}_{\text{inf},P_2} \right]^T,
\end{equation}
where $\textbf{r}_{\text{ref},p_1}$ is the $p_1^{th}$ reference signal, i.e.,
\begin{equation}
\textbf{r}_{\text{ref},p_1}=\left[r_{(p_1-1)L+1},\cdots,r_{p_1L}\right]^T ,\  p_1=1,\cdots,P_1.
\end{equation}
and $\textbf{r}_{\text{inf},p_2}$ is the $p_2^{th}$ information-bearing signal, i.e.,

\begin{equation}
\begin{aligned}
\textbf{r}_{\text{inf},p_2}= [ & r_{(P_1+p_2-1)L+1},r_{(P_1+p_2-1)L+2},
\\ & \cdots,r_{(P_1+p_2)L} ]^T, \  p_2=1,\cdots,P_2.
\end{aligned}
\end{equation}
The decision variable of the signal in the direct link with the $n^{th}$ Walsh code, $n\in\left\{ 1,2,\cdots,M \right\}$, can be derived as
\begin{align}
I_n=\sum^{P_1}_{p_1=1}[\textbf{r}_{\text{ref},p_1}]^T
\sum_{p_2=1}^{P_2}[w_{1+4(n-1),p_2} \textbf{r}_{\text{inf},p_2}].
\end{align}
The index symbol $a_l$ can be obtained by the code-index detection
\begin{align}
a_l=\text{arg}\max_{n=1,\cdots,M}(|I_n|)
\end{align}
Then, the modulated symbol can be obtained as
\begin{align}
b_l=\text{sign}(I_{a_l}).
\end{align}
In the backscatter link, $\textbf{R}'_l$ can be obtained as
\begin{equation}
 \textbf{R}'_l=\left[ \textbf{r}'_{\text{ref},1},\textbf{r}'_{\text{inf},1},
 \cdots, \textbf{r}'_{\text{ref},\delta},\textbf{r}'_{\text{inf},\delta} \right]^T,
 \ \delta=1,\cdots,(P_1+P_2)/4,
\end{equation}
where $\textbf{r}'_{\text{ref},\delta}$ is the ${\delta}^{th}$ reference signal, i.e.,
\begin{equation}
\begin{aligned}
\textbf{r}'_{\text{ref},\delta}= & [r_{(\delta-1)L+1},\cdots,r_{\delta L}]^T+
\\ & (-1)[r_{\delta L+1},\cdots,r_{(\delta+1)L}]^T,
\end{aligned}
\end{equation}
and $\textbf{r}'_{\text{inf},\delta}$ is the ${\delta}^{th}$ information-bearing signal, i.e.,
\begin{equation}
\begin{aligned}
\textbf{r}'_{\text{inf},\delta}= [r_{(\delta+1)L+1},\cdots,r_{(\delta+2)L}]^T-[r_{(\delta+2)L+1},\cdots,r_{(\delta+3)L}]^T.
\end{aligned}
\end{equation}
The decision variable of the $\delta^{th}$ symbol of the $l^{th}$ received signal $\textbf{R}'_l$ in the backscatter link is expressed by $ D_{l,\delta}=[\textbf{r}'_{\text{ref},\delta}]^T [\textbf{r}'_{\text{inf},\delta}]$. The transmitted symbol $b'_{l,\delta}$ in the backscatter link is yielded as $b'_{l,\delta}=\text{sign}(D_{l,\delta})$.
\section{Performance Analysis}
\subsection{BER Analysis of the Direct Link}
The average BER for the direct link of the CIM-DCSK-AmBC system consists of two parts: the BER of modulated bits (i.e., $P_{em}$) and the BER of index bits (i.e., $P_{eCIM}$), which can be expressed as
\begin{align}
P_{eCIM}=\frac{Q}{m_c}P_{ed}.
\end{align}
where $m_c=\log_2 M$ is the number of index bits per symbol, $Q$ is the expectation of the number of error index bits per symbol, i.e.,
\begin{align}
Q=\sum_{i=1}^{m_c}i\frac{ \left( \begin{matrix} m_c \\ i \end{matrix} \right) } {M-1},
\end{align}
where $\left( \begin{matrix} n \\ m \end{matrix} \right)=\frac{n!}{m!(n-m!)}$. $P_{ed}$ is the error probability of the code-index detection in the direct link of the CIM-DCSK-AmBC system, given as
\begin{align}
P_{ed}=\Pr \left( |I_{a_l}|<\max_{ n\neq a_l } \left \{ |I_n| \right \} \right),
\end{align}
where $I_n, n=1,5,\cdots,1+4(M-1)$ represent the direct link decision variables of the CIM-DCSK-AmBC system.
$P_{em}$ is expressed by
\begin{equation}
\label{Pem}
\begin{aligned}
P_{em}=P_e(1-P_{ed})+\frac{1}{2}P_{ed},
\end{aligned}
\end{equation}
Therefore, the average BER for the direct link is expressed by
\begin{equation}
\label{PeDirect}
\begin{aligned}
P_{e\text{Dir}}=\frac{m_c}{1+m_c}P_{eCIM}+\frac{1}{1+m_c}P_{em}.
\end{aligned}
\end{equation}

In the following, we derive the expression of $P_{ed}.$ Since $\textbf{S}_l$ is orthogonal to $\textbf{S}'_{l}$, $\textbf{S}'_{l}$ can be removed from the derivation of the average BER for the direct link. The correct decision variable $I_n (n=a_l)$ can be expressed as
\begin{equation}
\begin{aligned}
I_{n} &=\sum^{P_1}_{p_1=1}[\textbf{R}_{\text{ref},p_1}]^T \sum_{p_2=1}^{P_2}[w_{\hat n,p_2} \textbf{R}_{\text{inf},p_2}]
\\ & =\left[ P_1\textbf{x}^T + \sum^{P_1}_{p_1=1} \textbf{n}_{\text{ref},p_1}^T \right]\left[ P_2b_l\textbf{x}^T + \sum_{p_2=1}^{P_2} w_{n,p_2} \textbf{n}_{\text{inf},p_2}^T \right]^T,
\end{aligned}
\end{equation}
where $\textbf{n}_{\text{ref},p_1}^T$  and $\textbf{n}_{\text{inf},p_2}$ are the noise vectors with a mean of zero and a variance of $N_0/2$. Thus,
\begin{align}
E\{I_{n}\}=P_1P_2Lb_lE\{x_k^2\},
\end{align}
and
\begin{align}
Var\{I_{n}\}=(P_1^2 P_2 + P_1 P_1^2) L\frac{N_0}{2}E\{x_k^2\} + P_1P_2L\frac{N_0^2}{4}.
\end{align}

Similarly, the error decision variable $I_n (n\neq a_l)$ can be expressed as
\begin{equation}
\begin{aligned}
I_n &= \left[ P_1\textbf{x}^T+\sum^{P_1}_{p_1=1} \textbf{n}_{\text{ref},p_1}^T \right] \sum_{p_2=1}^{P_2} w_{n,p_2}\textbf{n}_{\text{inf},p_2}
\\ & = P_1\sum^{P_2}_{p_2=1}w_{n,p_2}\left[ \textbf{x}^T \textbf{n}_{\text{inf},p_2}  \right] + \sum^{P_1}_{p_1=1}\sum_{p_2=1}^{P_2} w_{n,p_2} \left[ \textbf{n}_{\text{ref},p_1}^T \textbf{n}_{\text{inf},p_2} \right].
\end{aligned}
\end{equation}
Hence,
\begin{align}
E\{I_n\}=0, Var\{I_n\}=P_1^2 P_2 L\frac{N_0}{2}E\{x_k^2\} + P_1P_2L\frac{N_0^2}{4}.
\end{align}
Substituting (21) into (22), (23), and (25), the means and variances are obtained as
\begin{align}
\mu_1=\frac{P_1P_2}{P_1+P_2}E_s, \quad \mu_2=0,
\end{align}
%\begin{equation}
%\begin{aligned}
%\sigma_1^2 & =\frac{1}{2} P_1 P_2 E_s N_0 + \frac{1}{4} P_1 P_2 L N_0^2
%\\ & =E_sN_0\left( \frac{P_1 P_2}{2} + \frac{P_1P_2L}{4\gamma_s}  \right),
%\end{aligned}
%\end{equation}
\begin{equation}
\begin{aligned}
\sigma_1^2 \!=\!\frac{1}{2} P_1 P_2 E_s N_0 \!+\! \frac{1}{4} P_1 P_2 L N_0^2
 \!=\!E_sN_0\left( \frac{P_1 P_2}{2} \!+\! \frac{P_1P_2L}{4\gamma_s}  \right)\!,
\end{aligned}
\end{equation}
\begin{equation}
\begin{aligned}
\sigma_2^2 & =\frac{1}{2}\frac{P_1^2 P_2}{P_1+P_2} E_sN_0 + \frac{1}{4} P_1P_2 L N_0^2
\\ & =E_sN_0 \underbrace{ \left( \frac{P_1^2 P_2}{2(P_1+P_2)} + \frac{P_1P_2L}{4\gamma_s}  \right) }
\\ & \qquad \qquad \qquad \qquad =\Omega
\end{aligned},
\end{equation}
where $E_s=\left(P_1+P_2\right)E\left\{x_k^2\right\}$ is the energy of signal per symbol, and $\gamma_s=E_s/N_0$ is the SNR of the CIM-DCSK-AmBC system.
Let $X_1 = \text{max} \{|I_n|\} , n \neq a_l$, $Y_1=|I_{a_l}|$. Afterwards, $P_{ed}$ can be derived as
\begin{equation}
\label{Ped}
\begin{aligned}
P_{ed} & =\Pr \{Y_1<X_1\}
\\ & =\int_{0}^{\infty}\Pr \{ y<X_1 \}f_{Y_1}(y)dy
\\ =\frac{1}{\sqrt{2\pi\rho }} & \int_{0}^{\infty}  \left[ 1-\left[ \text{erf} \left( \frac{u}{2\Omega} \right)^{M-1} \right] \right] \{ e^{-\frac{(u-\Gamma)^2}{2\rho}} + e^{-\frac{(u+\Gamma)^2}{2\rho}} \}du,
\end{aligned}
\end{equation}
where
\begin{equation}
\begin{aligned}
\Gamma = & \sqrt{ \frac{P_1P_2+\frac{1}{2}P_1P_2L/\gamma_s}{\pi} }
e^{-\frac{ \frac{P_1P_2}{(P_1+P_2)^2}\gamma_s^2 } {\gamma_s+\frac{1}{2}L} }
\\ & -\frac{P_1P_2}{P_1+P_2}\sqrt{\gamma_s}
\text{erf}(-\frac{ \frac{P_1P_2}{(P_1+P_2)^2}\gamma_s^2 } {\gamma_s+\frac{1}{2}L}),
\end{aligned}
\end{equation}
\begin{equation}
\begin{aligned}
\rho = (\frac{P_1+P_2}{P_1P_2})^2\gamma_s + \frac{1}{2}P_1P_2 + \frac{P_1P_2L}{4\gamma_s} - \Gamma^2,
\end{aligned}
\end{equation}

Besides, $P_e$ can be derived as
\begin{equation}
\label{Pe}
\begin{aligned}
P_{e} & = \frac{1}{2}\text{erfc}\left( \left[ \frac{2Var\{I_{a_l}\}}{E\{I_{a_l}\}^2} \right]^{-0.5} \right)
\\ & = \frac{1}{2}\text{erfc} \left(  \left[ \frac{\left(P_1+P_2\right)^2}{P_1P_2\gamma_s} + \frac{(P_1+P_2)\beta}{2P_1P_2\gamma_s^2} \right]^{-0.5} \right),
\end{aligned}
\end{equation}
where $\beta=(P_1+P_2)L$ is the spreading factor.

Finally, $P_{e\text{Dir}}$ can be obtained by substituting (\ref{Ped}), (\ref{Pe}), and (\ref{Pem}) into (\ref{PeDirect}).
\vspace{-0.45cm}
\subsection{BER Analysis of the Backscatter Link}
Similar to the derivation of the average BER for the direct link, $\textbf{S}_l$ can be removed from the derivation of the backscatter link BER, thus the decision variable in the backscatter link is expressed by
\begin{equation}
\begin{aligned}
D'_{l,\delta} & =[\textbf{r}'_{\text{ref},\delta}] [\textbf{r}'_{\text{inf},\delta}] ^T
\\ & =\left[ 2\textbf{x}^T + \sum_{p_1=1}^2 \textbf{n}'_{\delta,p_1} \right] \left[ 2\textbf{x}^T + \sum_{p_2=3}^4 \textbf{n}'_{\delta,p_2} \right]^T,
\end{aligned}
\end{equation}
where $\textbf{n}'_{\delta,p}=[n_{(\delta+p-1)L+1},\cdots,n_{(\delta+p)L}]$ is additive white Gaussian noise vector.

Then, the mean and variance of $D'_{l,\delta}$ are respectively given by $E\{ D'_{l,\delta} \}=4LE\{x_k^2\}$, and $Var\{ D'_{l,\delta} \}=8LN_0E\{x_k^2\} + L N_0^2$.

Consequently, the BER of the backscatter link can be derived as
\begin{align}
P_{e\text{BC}}=\frac{1}{2}\text{erfc}\left( \left[ \frac{P_1+P_2}{\gamma_s} + \frac{(P_1+P_2)\beta}{8\gamma_s^2} \right]^{-0.5} \right),
\end{align}
where $\beta=(P_1+P_2)L$ is the spreading factor.
\vspace{-0.45cm}
\subsection{Multipath Rayleigh fading channel}
We assume that the path delay $\tau$ satisfies $\tau \ll \beta$. In this case, the inter-symbol interference can be ignored. For the direct link, we define channel 1 is an $L_h$ paths Rayleigh slow fading channels with same power gain in each channel. We define $E_b=E_s/N_{bits}$, $N_{bits}$ is the number of the total bits transmitted in the direct link and the backscatter link of the system. The PDF of the Rayleigh fading channels is given by
\begin{align}
f_{\text{Ray}}(\gamma_h)=\frac{\gamma_h^{L-1}}{(L-1)!\bar{\gamma}^L}e^{-\frac{\gamma_h}{\gamma}},
\end{align}
where $\gamma_h=\sum_{l=1}^{L}{h_l^2E_b/N_0}$, and $\bar{\gamma}=(1/L)E_b/N_0$. The BER for the direct link of the proposed system over multipath Rayleigh fading channel can be calculated by
\begin{align}
\text{BER}_\text{Dir}=\int_0^{+\infty}{P_{e\text{Dir}}f_{\text{Ray}}(\gamma_h)d\gamma_h}.
\end{align}

For the backscatter link, we consider a cascade channel with two multipath Rayleigh fading channels channel 2 and channel 3. $L_f$ and $L_g$ are the number of paths for channel 2 and 3, respectively. We assume $E[f_1^2]=\cdots=E[f_{L_f}^2]$ and $E[g_1^2]=\cdots=E[g_{L_g}^2]$. Path loss coefficient of channel 2 and channel 3 are set to $\xi_f$ and $\xi_g$, respectively. Thus, we set $\sum_{l=1}^{L_f}{E[f_l^2]}=\xi_f$ and $\sum_{l=1}^{L_g}{E[g_l^2]}=\xi_g$ for channel 2 and 3, respectively. The BER for the backscatter link of the proposed system over multipath Rayleigh fading channel can be calculated by
\begin{align}
\text{BER}_\text{BC}=\xi_f\xi_g\int_0^{+\infty}\int_0^{+\infty}{f_\chi(z){P_{e\text{BC}}f_{\text{Ray}}(\gamma_f)d\gamma_f}dz},
\end{align}
where $f_\chi(z)=\frac{1}{2^(k/2)\Gamma(k/2)}z^{k/2-1}e^{-z/2}$ is the PDF of chi-square distribution, $k=2L_f$.

\subsection{Throughput Analysis of the Backscatter Link}

The normalized throughput in a wireless communication system is defined as the ratio between the successfully received bits/transmission period and the transmitted bits/transmission period\cite{1246003}. The transmission period $\bar{Q}$ is defined as the transmission duration for each packet in a CIM-DCSK-AmBC system. That means $\bar{Q}=\left( N_p/N_{\text{CIM-BC}} \right) T_d$, where $N_p$ is the number of transmitted bits in each packet, $N_{\text{CIM-BC}}=\left( P_1+P_2 \right)/4$ is the number of bits transmitted by the backscatter link of the CIM-DCSK-AmBC system in each symbol, and $T_d$ is the duration for each symbol. The normalized throughput for the backscatter link of a DCSK based AmBC system is given by
\begin{align}
R_t=\left[ \left( 1-\text{BER}_{t} \right)^N_p\bar{Q} \right]/\bar{Q_t},
\end{align}
where $\bar{Q_t}$ is the transmission period for the given system $t\in \left\{ \text{CIM-BC},\text{SR-BC} \right\}$. Consequently, the transmission periods for the backscatter link of the CIM-DCSK-AmBC system and the SR-DCSK-AmBC system are $\bar{Q}_\text{CIM}=\left( N_p/N_{\text{CIM-BC}} \right) T_d$ and $\bar{Q}_\text{SR}=\left( N_p/1 \right) T_d$, respectively. As a result, the normalized throughputs of the aforementioned systems become $R_\text{CIM}=\left( 1-\text{BER}_{\text{CIM-BC}} \right)^{N_p}$ and $R_\text{SR}=\left( 1-\text{BER}_{\text{SR-BC}} \right)^{N_p}/N_{\text{CIM-BC}}$, where $\text{BER}_{\text{CIM-BC}}$ is the BER for backscatter link of the CIM-DCSK-AmBC system, and $\text{BER}_{\text{SR-BC}}$ is the BER for backscatterlink of the SR-DCSK-AmBC system.
\vspace{-0.2cm}
\section{Numerical Results and Discussions}
In this section, simulation results are carried out to evaluate the performance of the proposed CIM-DCSK-AmBC system over multipath Rayleigh fading channels. In all figures, for multipath Rayleigh fading channels $h$ and $f$, which are defined in Fig.~1, the number of paths is set as $L_h=L_f=2$, with different time delays $\tau_1=0$, $\tau_2=3$. We set path loss $\xi_f=0.7$ and $\xi_g=0.6$. The average power gains of $h$ and $f$ are $E[h_1^2]=E[h_2^2]=1/2$ and $E[f_1^2]=E[f_2^2]=1/2\xi_f=1/2*0.7$, respectively, where $*$ is defined as the multiplication sign. For the multipath Rayleigh fading channel $g$, which is defined in Fig.1, the number of paths is set as $L_g=3$, with equal average power gains $E[g_1^2]=E[g_2^2]=E[g_3^2]=1/3\xi_g=1/3*0.6$, and different time delays $\tau'_1=0$, $\tau'_2=1$, $\tau'_3=2$.
\begin{figure}[tbp]
\center
\vspace{-0.0cm}
\includegraphics[width=2.5 in]{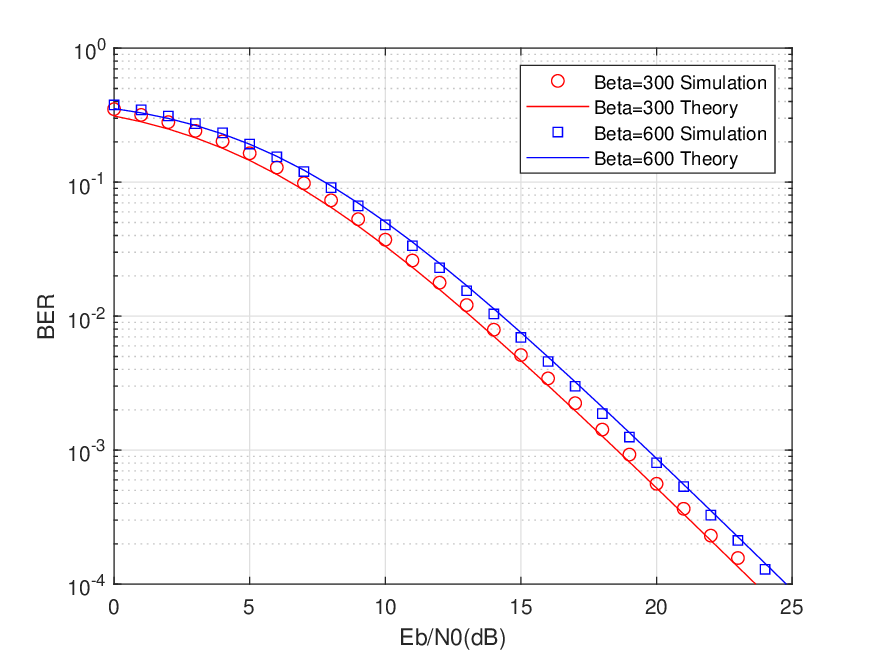}
\vspace{-0.3cm}
\caption{Simulated and analytical BERs in the direct link of the proposed CIM-DCSK-AmBC system over multipath Rayleigh fading channel for different $\beta$ with $P_1=4$, $P_2=8$, $\tau_1=0$, $\tau_2=3$, $\zeta=1$.}
\label{DirBER}
\vspace{-0.4cm}
\end{figure}
\begin{figure}[tbp]
\center
\vspace{-0.0cm}
\includegraphics[width=2.5 in]{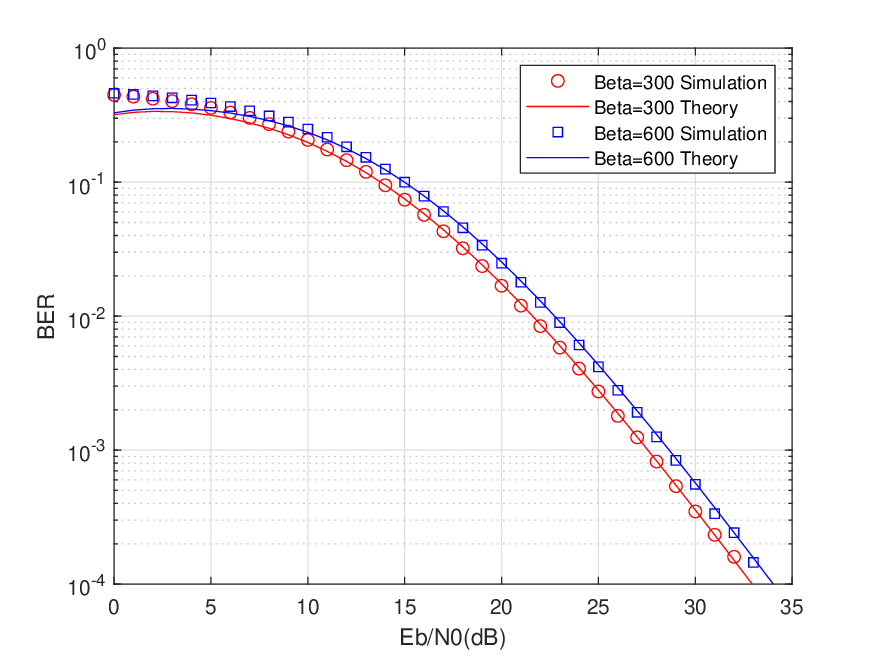}
\vspace{-0.3cm}
\caption{Simulated and analytical BERs in the backscatter link of the proposed CIM-DCSK-AmBC system over multipath Rayleigh fading channel for different $\beta$ with $P_1=4$, $P_2=8$, $\tau_1=0$, $\tau_2=3$, $\tau'_1=0$, $\tau'_2=1$, $\tau'_3=2$, $\zeta=1$.}
\label{BSBER}
\vspace{-0.7cm}
\end{figure}

Fig.~\ref{DirBER} and \ref{BSBER} show the theoretical and simulated BER results of the proposed CIM-DCSK-AmBC system over multipath Rayleigh fading channels. We assume that the order of Walsh code is $P_2=8$. It shows that the simulated results are in good agreement with the theoretical results. Thereby, we can conclude the interference between the direct link and the backscatter link has been well eliminated.

\begin{figure}[tbp]
\center
\vspace{-0.0cm}
\includegraphics[width=2.5 in]{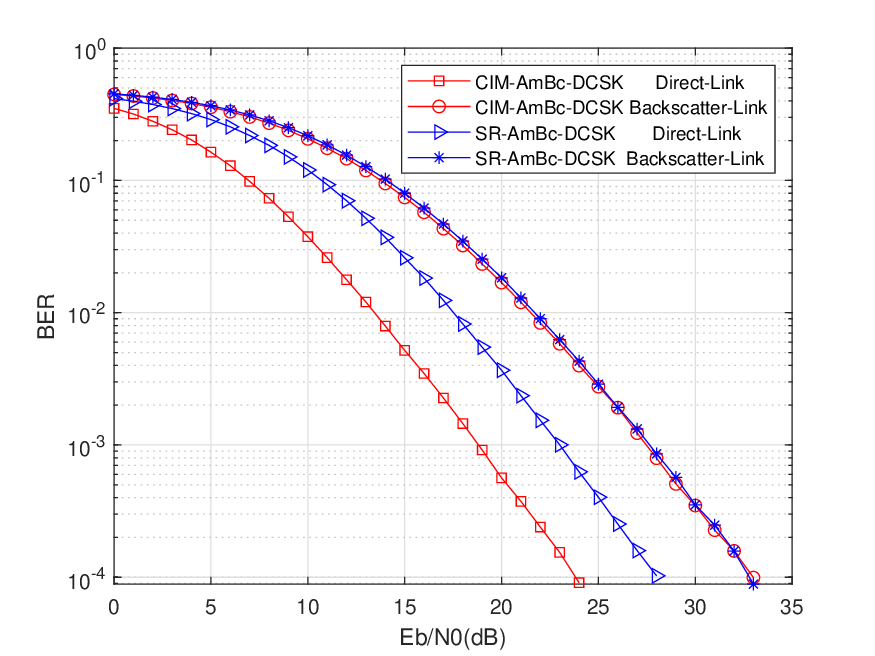}
\vspace{-0.3cm}
\caption{BER performance comparison between the CIM-DCSK-AmBC and the SR-DCSK-AmBC systems over a multipath Rayleigh fading channel with $\beta=300$, $P_1=4$, $P_2=8$, $\zeta=1$ for the CIM-DCSK-AmBC and $\beta=297$, $P=8$, $\zeta=1$ for the SR-DCSK-AmBC.}
\label{BERComparison}
\vspace{-0.4cm}
\end{figure}

\begin{figure}[tbp]
\center
\vspace{-0.0cm}
\includegraphics[width=2.5 in]{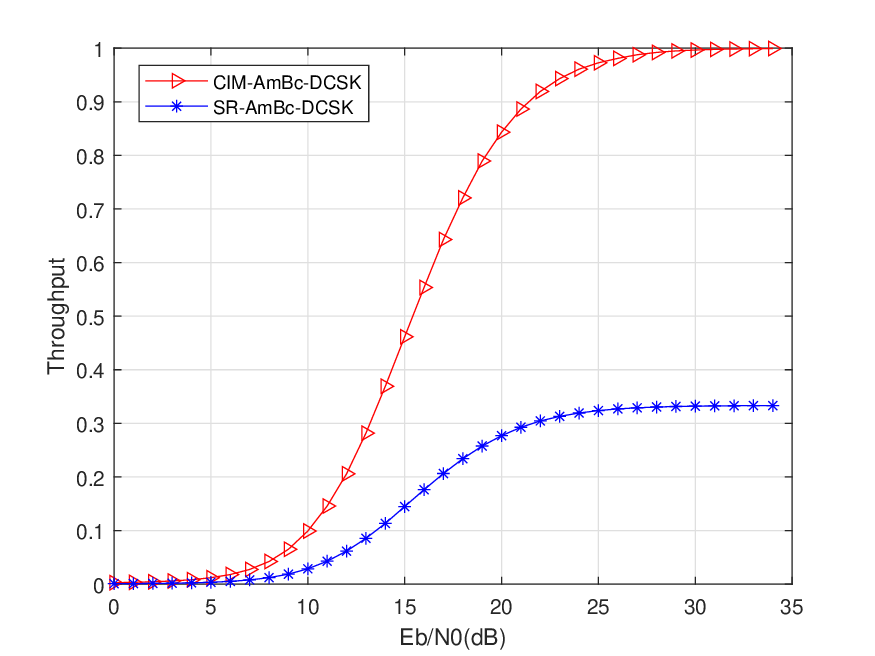}
\vspace{-0.3cm}
\caption{Backscatter-link throughput comparison between the CIM-DCSK-AmBC and the SR-DCSK-AmBC systems over a multipath Rayleigh fading channel with $N_p=12$, $\beta=300$, $P_1=4$, $P_2=8$, $\zeta=1$ for the CIM-DCSK-AmBC and $N_p=12$, $\beta=297$, $P=8$, $\zeta=1$ for the SR-DCSK-AmBC.}
\label{BSThrouputComparison}
\vspace{-0.75cm}
\end{figure}

The BER comparison of the CIM-DCSK-AmBC and the SR-DCSK-AmBC systems is shown in Fig.~\ref{BERComparison}. For a fair comparison, the spreading factor of the CIM-DCSK-AmBC system is set to $\beta=300$ with $P_1=4$ and $P_2=8$, while the spreading factor of the SR-DCSK-AmBC system is set to $\beta=297$ with $P=8$. It is because the spreading factor of the SR-DCSK-AmBC system should be divisible by $9$ and cannot be set to $300$. As shown in this figure, there is a distinct BER advantage of the CIM-DCSK-AmBC system over the SR-DCSK-AmBC system in the direct link. To be more specific, the gain of the CIM-DCSK-AmBC system is about 4.5 dB at a BER of $10^{-4}$ with respect to the SR-DCSK-AmBC system over the multipath Rayleigh fading channel. The BER performance of the CIM-DCSK-AmBC system is almost the same as that of the SR-DCSK-AmBC system in the backscatter link. Nevertheless, it can be observed in Fig.~\ref{BSThrouputComparison} that the throughput for the backscatter link of the CIM-DCSK-AmBC is three times that of the SR-DCSK-AmBC. This observation is due to the fact that the proposed system designs the signal format particularly for the backscatter link, thus transmitting additonal information bits.

\begin{figure}[tbp]
\center
\vspace{-0.0cm}
\includegraphics[width=2.5 in]{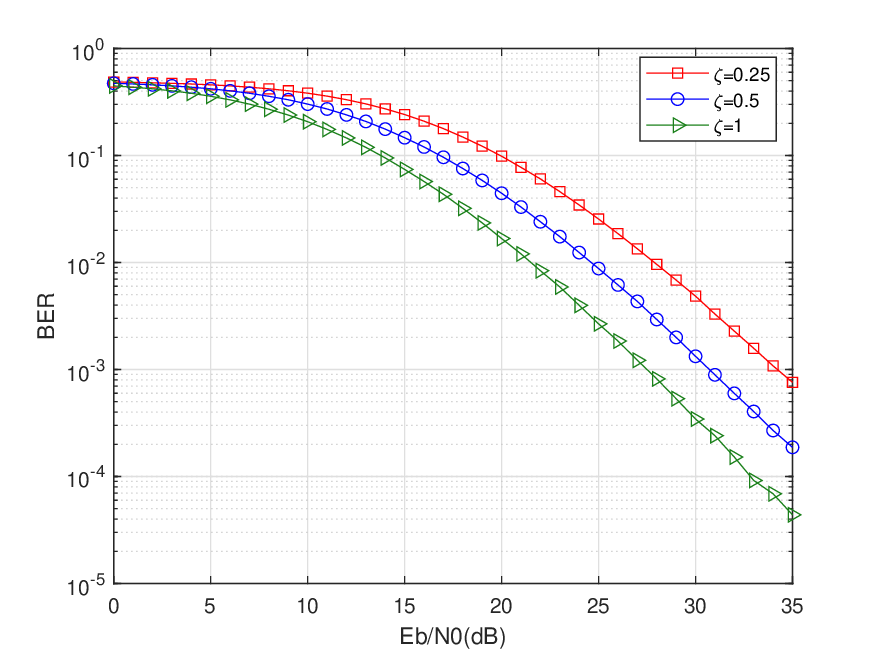}
\vspace{-0.3cm}
\caption{Simulated BERs in the backscatter link of the CIM-DCSK-AmBC over a multipath Rayleigh fading channel for different $\zeta$ with $\beta=300$, $P_1=4$, $P_2=8$.}
\label{Zeta}
\vspace{-0.75cm}
\end{figure}

As a further study, the BER performance of the CIM-DCSK-AmBC system over different reflecting coefficients $\zeta$ is shown in Fig.~\ref{Zeta}. As seem, with the decrease of the reflecting coefficient $\zeta$, the BER performance gradually deteriorates. This is because the decrease of $\zeta$ leads to lower reflecting energy in the backscatter link.
\vspace{-0.25cm}
\section{Conclusion}
In this paper, a novel CIM-DCSK-AmBC system has been proposed to improve the data rate of the conventional DCSK-AmBC system. In the proposed system, the Walsh code used in the direct link signal has been intelligently selected to adapt to the signal format in the backscatter link, which allows the receiver to eliminate the interference between the two links. The signal format in the backscatter link has been designed to transmit additional information bits, while maintaining the anti-direct link interference capability. The BER expressions of the CIM-DCSK-AmBC system have been derived and verified over multipath Rayleigh fading channel. Simulation results demonstrate that the BER performance in the direct link of the proposed system is much better than that of the SR-DCSK-AmBC system. Moreover, the throughput in the direct link of the proposed system is remarkably higher than that of the SR-DCSK-AmBC system without any BER degradation. Owing to the above benefits, the proposed CIM-DCSK-AmBC system can be considered a competitive candidate for low-power and high-data-rate short-range wireless communications.
\vspace{-0.25cm}
%\bibliographystyle{IEEEtran}
%\bibliography{Reference}

\end{document}